\documentclass[10pt,twocolumn]{ieeetran} 
\usepackage{times}
\usepackage{graphicx}

\newcommand{\figuradue}[3]
{
\begin{figure}
  \centering
 \includegraphics[width=8cm]{#1} %
  \caption{#2}\label{#3}
\end{figure}
}

\newcommand{\figuramedia}[3]
{
\begin{figure*}
  \centering
 \includegraphics[width=15cm]{#1}
  \caption{#2}\label{#3}
\end{figure*}
}

\newcommand{\figuramatlab}[3]
{
\begin{figure}
  \centering
 \includegraphics[width=7.5cm]{#1} 
  \caption{#2}\label{#3}
\end{figure}
}
\newcommand{\figuragrossa}[3]
{
\begin{figure}
  \centering
 \includegraphics[width=7.5cm]{#1}
  \caption{#2}\label{#3}
\end{figure}
}

\begin{document}
\title{Fixed-Point Design of Generalized Comb
Filters: A Statistical Approach}
\author{Massimiliano
Laddomada,~\IEEEmembership{Member,~IEEE}\\
\thanks{Massimiliano Laddomada is with the Electrical Engineering Department of Texas A\&M University-Texarkana, USA. E-mail: \textrm{mladdomada@tamut.edu}.}}
\maketitle
\begin{abstract}
This paper is concerned with the problem of designing
computationally efficient Generalized Comb Filters (GCF).
Basically, GCF filters are anti-aliasing filters that guarantee
superior performance in terms of selectivity and quantization
noise rejection compared to classical comb filters, when used as
decimation filters in multistage architectures.

Upon employing a partial polyphase (PP) architecture proposed in a
companion paper, we develop a sensitivity analysis in order to
investigate the effects of the coefficients' quantization on the
frequency response of the designed filters.

We show that the sensitivity of the filter response to errors in
the coefficients is dependent on the particular split of the
decimation factor between the two sub-filters constituting the PP
architecture. The sensitivity analysis is then used for developing
a fixed-point implementation of a sample filter from the class of
GCF filters, used as reference filter throughout the paper.

Finally, we present computer simulations in order to evaluate the
performance of the designed fixed-point filters.
\end{abstract}
\section{Introduction and problem formulation}
The design of computationally efficient decimation filters for
oversampled $\Sigma\Delta$ A/D
converters~\cite{Temes}-\cite{candy_decim}, as well as for
classical oversampled A/D converters, has received a renewed
interest recently, spurred by intense research activities in
connection to the design of digital front-ends for both wideband
digital receivers and Software Defined Radio
receivers~\cite{Mitola}-\cite{Laddomada_transceiver}.

Consider a base-band analog signal $x(t)$ (with bandwidth
$\left[-f_x,+f_x\right]$) sampled by an Analog-to-Digital (A/D)
converter at rate $f_s=1/T_s=2\rho f_x\gg 2f_x$, where $\rho$($\ge
1$) is the so-called oversampling ratio. If $\rho$ is close to
unity, the A/D converter operates at the Nyquist frequency,
whereas for $\rho\gg 1$ we are referring to oversampled A/D
converter. When $\rho\gg 1$, the decimation of the oversampled
discrete-time signal $x(n/f_s)$ is usually accomplished by
cascading two (or more) decimation stages, followed by a FIR
filter that provides the required selectivity on the discrete-time
signal $x(nT_N)$ at baseband.

Fig.~\ref{arch} shows a multistage decimation architecture
composed by $m$ decimation stages operating on the oversampled
signal $x(nT_s)$. Also shown are the data rates of the sampled
data at the input, as well as at the output, of each decimation
stage in the multistage decimation architecture.

For the sake to contain the computational complexity of the
overall architecture, the first decimation stage usually employs a
multiplier-less filter~\cite{CrochiereRabiner}-\cite{dolecek_6}. A
widely used filter featuring this property is the comb
filter~\cite{Hoge,candy_decim,Chu}, which provides an intrinsic
anti-aliasing effect by placing its zeros in the middle of each
folding band, i.e., in the integer multiples of the digital
frequency $1/D$ ($D$ is the decimation factor). The transfer
function of a $N_c$-th-order comb filter is defined
as~\cite{Hoge}:
\figuramedia{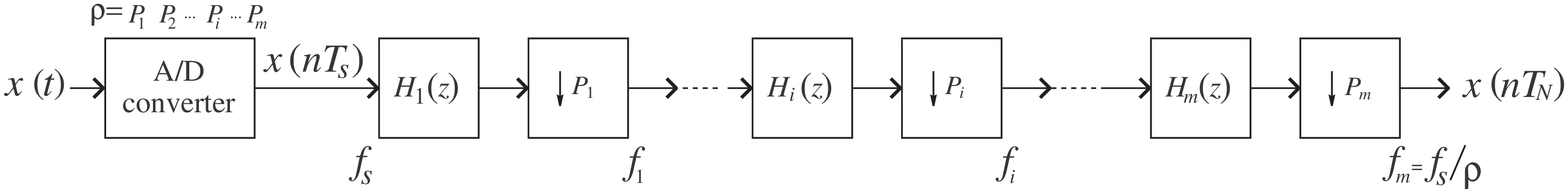}{General architecture of
a multi-stage decimation architecture for oversampled A/D
converters. Given $\rho$, the discrete-time signal $x(n/f_s)$ is
decimated through a cascade of $m$ decimation stages, obtaining
the Nyquist-sampled discrete-time signal $x(nT_N)$ at baseband.
The $i$-th decimation stage is composed by an anti-aliasing filter
$H_i(z)$ followed by a decimator by $P_i$. The filter $H_i(z)$
operates at the data rate $f_{i-1}$, while the $i$-th decimation
stage delivers a discrete-time signal with rate $f_i=f_{i-1}/P_i$.
The index $i$ takes on values in the range
$\{1,\ldots,m\}$.}{arch}
\begin{equation}\small
H_{C}(z)=\left(\frac{1}{D}\frac{1-z^{-D}}{1-z^{-1}}\right)^{N_c}=
\frac{1}{D^{N_c}}\prod_{i=1}^{D-1}\left(1-z^{-1}e^{j\frac{2\pi}{D}i}\right)^{N_c},
\label{transf_funct_CIC_N}
\end{equation}
where $D$ is the decimation factor.

Unless the design of classical FIR filters, the design of a
decimation filter embedded in a multistage architecture imposes
stringent constraints on proper bandwidths. Let us elaborate.

Consider the architecture shown in Fig.~\ref{arch}, where the
oversampling ratio $\rho$ is factorized as
\[
\rho=\prod_{i=1}^{m}P_i.
\]
In the previous relation, each $P_i$ is a proper positive integer.
The sampling rates at the input and output of the $i$-th stage
are, respectively,
\[
f_{i-1}=f_{i}\cdot P_{i},~\forall i=1,\ldots,m,
\]
with $f_0=f_s$, and
\[
f_i=\frac{f_s}{\prod_{k=1}^{i}P_k},~\forall i=1,\ldots,m.
\]
With this setup, consider the frequency response $H_i(f_d)$ of the
$i$-th decimation filter pictorially shown in
Fig.~\ref{pictorial_frequency_response}. The digital frequency
$f^{i-1}_c$ is the normalized signal bandwidth at the input of the
$i$-th decimation filter. Notice that, for $i=1$, it is
$f^o_c=f_x/f_s=(2\rho)^{-1}$: this is the normalized bandwidth of
the signal sampled by the A/D converter at rate $f_s$. For any
other $i$, the relation between $f^i_c$ and $f^o_c$ is
\[
f^i_c=f^{i-1}_c P_{i},~\forall i=1,\ldots,m.
\]
Given this setup, the frequency response $H_i(f_d)$ has to
attenuate the quantization noise (QN) within the frequency bands
\begin{equation}\label{folding_bands_def}
\begin{array}{lll}
\left[\frac{k}{P_i}-f^{i-1}_c;\frac{k}{P_i}+f^{i-1}_c\right],&k=1,...,k_M&\\
k_M=\lfloor \frac{P_i}{2}\rfloor,&P_i~\textrm{even}&\\
k_M=\lfloor \frac{P_i-1}{2}\rfloor,&P_i~\textrm{odd}&
\end{array}
\end{equation}
because the QN falling inside these frequency bands, will fold
down to baseband (i.e., within the useful signal bandwidth
$\left[-f^{i-1}_c,+f^{i-1}_c\right]$) due to the sampling rate
reduction by $P_i$ in the $i$-th decimation stage
\cite{laddomada_multistage_dec}. Such a QN will irremediably
affect the signal resolution after the multistage decimation
architecture. On the other hand, the frequency ranges labeled as
\textit{don't care} bands in
Fig.~\ref{pictorial_frequency_response}, do not require stringent
selectivity, since the QN within these bands will be rejected by
the subsequent anti-aliasing filters in the multistage chain.

With this background, let us provide a survey of the recent
literature related to the problem addressed in this paper.
Tutorials on the design of multirate filters can be found
in~\cite{CrochiereRabinerTut,Vaidyanathan}, while essential books
on this topic are~\cite{CrochiereRabiner}-\cite{dolecek_6}. The
design of optimized multistage decimation and interpolation
filters has been recently addressed by Coffey
in~\cite{Coffey1}-\cite{Coffey2}, while the design of multistage
decimation architectures relying on constituent cyclotomic
polynomial filters has been presented in
\cite{laddomada_multistage_dec}. A $3$-rd-order modified
decimation sinc filter was proposed in~\cite{Letizia}, and
developed in~\cite{max}. The class of comb filters was generalized
in~\cite{laddomada_gcf}, whereby the author proposed an
optimization framework for deriving the optimal zero rotations of
GCFs for any filter order and decimation factor $D$.

Other works somewhat related to the topic addressed in this paper
are~\cite{gao}-\cite{laddomada_sharp}. In~\cite{gao}
and~\cite{aboushady}, the authors proposed computational efficient
decimation filter architectures for implementing non recursive
classical comb filters. In~\cite{kwentus}, the authors proposed
the use of decimation sharpened filters embedding comb filters,
whereas in~\cite{dolecek}-\cite{dolecek_2} the authors addressed
the design of a novel two-stage sharpened comb decimator.
In~\cite{laddomada}, the authors proposed novel decimation schemes
for $\Sigma\Delta$ A/D converters based on Kaiser and Hamming
sharpened filters, then generalized in \cite{laddomada_sharp} for
higher order decimation filters. Papers
\cite{dolecek_3}-\cite{dolecek_5} focus on the design of
decimation filters with improved frequency responses.

The main aim of this paper is to develop a mathematical framework
for the design of fixed-point GCF decimation filters relying on a
partial polyphase FIR architecture proposed in the companion
paper~\cite{laddomada_part_poly}. To this end, we develop a
sensitivity analysis in order to investigate the effects of the
coefficients' quantization on the frequency response of the
designed filters.

For conciseness, we focus on the design of the first decimation
filter in the multistage architecture in Fig.~\ref{arch}, even
though the proposed analysis can be easily extended to the design
of the other anti-aliasing filters in the cascade.

The sensitivity analysis paves the way to a statistical approach
useful to identify the coefficient word lengths of the proposed
fixed-point architecture. Moreover, we show that the proposed
analysis highlights some key issues in connection to the choice of
the proper split of the decimation factor between the polyphase
stage and the cascaded FIR sections of the employed partial
polyphase architecture.

The rest of the paper is organized as follows. In
Section~\ref{overview_section}, we briefly review the transfer
functions of GCF filters, as well as the partial polyphase
architecture employed throughout the paper, and outline the key
advantages that these filters feature with respect to classical
comb filters. Section~\ref{Sensitivity_Analysis_section} presents
a mathematical framework for evaluating the sensitivity of the
frequency response of GCF filters to the quantization of the
coefficients. In
Section~\ref{Comparisons_Simulation_Results_Section}, we discuss
general guidelines for the design of the proposed filters, and
present some simulation results. Finally,
Section~\ref{conclusions} draws the conclusions.
\figuradue{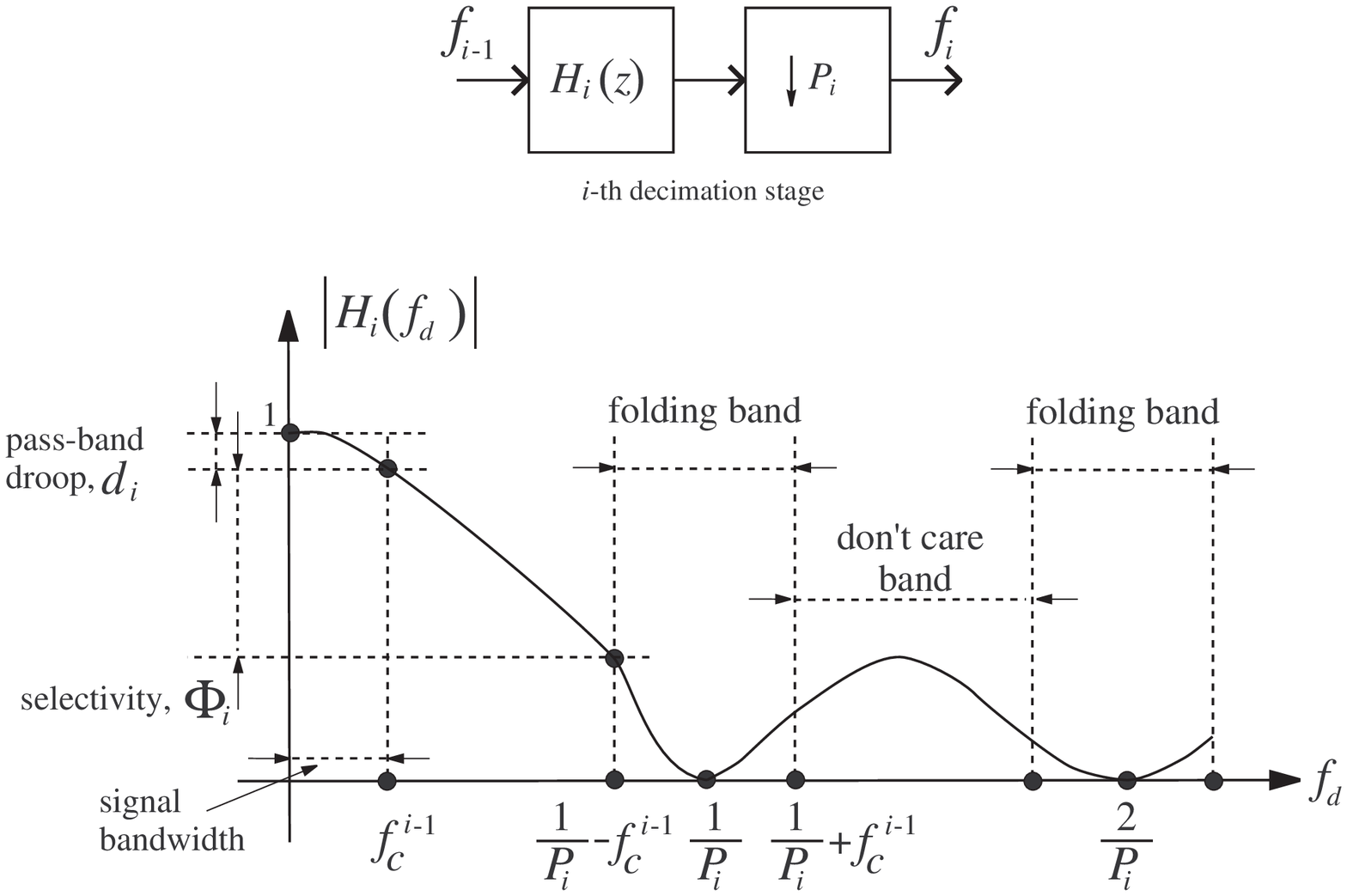}{Pictorial
representation of the frequency response of the $i$-th decimation
filter, $H_i(z)$, along with the key frequency intervals to be
carefully considered during the design. $P_i$ is the decimation
factor, $f_{i-1}$ is the data rate at the input, $f_i$ is the data
rate of the decimated data, and $f_c^{i-1}$ is the normalized
signal bandwidth.}{pictorial_frequency_response}
\section{Overview of GCF filters: The Partial Polyphase Architecture}
\label{overview_section}
\figuramedia{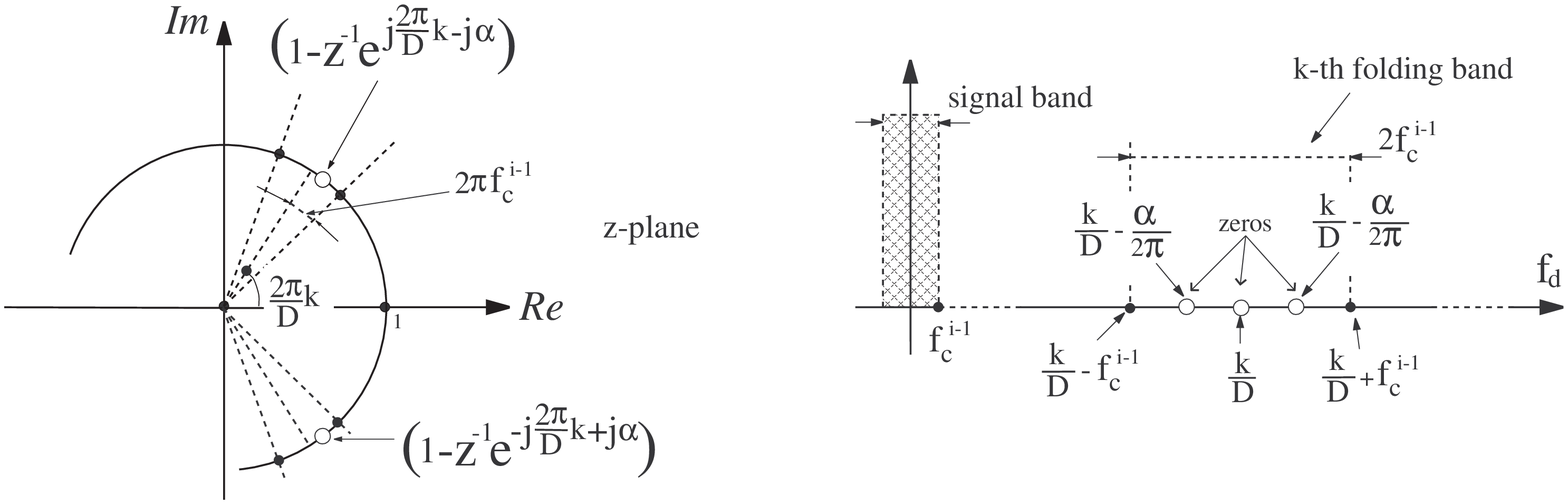}{Zero locations of GCF filters
within the $k$-th folding band. Zeros are displayed in the
$z$-plane, as well as in the frequency domain, in order to
highlight the main differences between GCF and classical comb
filters.}{zeros_posi_GCF}
Our objective in this section is to recall the fundamental
concepts for analyzing both the time-domain behaviour and the
frequency response of GCF filters, as well as to highlight the
main differences between GCF and classical comb filters.

For conciseness, our discussion will be restricted to a 3rd-order
GCF filter, which will be used as a reference scheme throughout
the paper, and we will present the non-recursive, partial
polyphase architecture\footnote{Even though both recursive and non
recursive implementations can be devised for GCF filters, the
non-recursive architecture does not present any instability
problem deriving from the quantization of the coefficients. We
notice in passing that recursive GCF filter realizations rely on
zero-pole cancellations, which can be impaired by the quantization
of the coefficients. This is the basic reason for the use of this
FIR architecture in the developments that follow.} developed in
the companion paper \cite{laddomada_part_poly}.

\noindent Let us focus on the design of the $i$-th decimation
filter, $H_i(z)$, in Fig.~\ref{arch} and, for ease of notation,
assume $P_i=D$ and omit the subscript $i$ in $H_i(z)$. Given $D$,
and recalling the definition of the folding bands
in~(\ref{folding_bands_def}), a classical 3rd-order comb filter
(see~(\ref{transf_funct_CIC_N}) with $N_c=3$) presents
$3$-rd-order zeros in the complex locations
$$z_k=e^{j\frac{2\pi}{D}k},~\forall k=1,\ldots,D-1,$$
or, equivalently, in the digital frequencies
$f_{z_k}=\frac{k}{D},~k\in \{1,\ldots,k_M\}$. Therefore, a
3rd-order zero is placed in the middle of each folding band. This
idea is illustrated in Fig.~\ref{zeros_posi_GCF}, where the $k$-th
folding band is shown in the $z$-plane, as well as in the
frequency domain.

On the other hand, a 3rd-order GCF filter places, in the $k$-th
folding band, $3$ zeros in the digital frequencies
$\frac{k}{D}-\frac{\alpha}{2\pi}, \frac{k}{D}$, and
$\frac{k}{D}+\frac{\alpha}{2\pi}$, whereas the edges of the $k$-th
folding band are $\frac{k}{D}-f_c^{i-1}$ and
$\frac{k}{D}+f_c^{i-1}$. Therefore, as shown in
Fig.~\ref{zeros_posi_GCF}, the choice $\alpha=q 2\pi
f_c^{i-1},$~with $q\in \left[0,+1\right]$, allows a better
distribution of the three zeros within the $k$-th folding band,
whose width is strictly related to the bandwidth $f_c^{i-1}$ of
the useful discrete-time signal.

The optimal parameter $q$ has been found in \cite{laddomada_gcf},
and we will employ such a value throughout this work. As an
example, the optimal value $q=0.79$ is such that a 3rd-order GCF
filter features an additional 8dB of QN rejection over a classical
3rd-order comb filter around the folding bands.

Once again, let us focus our attention on the 3rd-order GCF
filter, and consider a decimation factor $D$ that can be expressed
as the $p$-th power-of-two, i.e., $D=2^p$, where $p$ is a suitable
integer greater than zero. Moreover, let us factorize the
decimation factor $D$ as $D=D_1\cdot D_2$, whereby
$D_1=2^{p_p+1}$, $D_2=2^{p-p_p-1}$, and $p_p$ can take on any
integer value in the set $\left\{-1,\ldots,p-1\right\}$. With this
setup, the $z$-transfer function of a third-order GCF filter
realized with the partial polyphase FIR architecture in
Fig.~\ref{n_rec_impl_gcf3}, can be defined as follows:
\begin{eqnarray}\label{third_order_gcf3_2_2}
H(z) &=&H_o\cdot H_{P}(z) \cdot H_{N}(z),
\end{eqnarray}
whereby $H_o$ is a constant term ensuring unity gain at
baseband\footnote{For simplicity, we omit this constant term in
the derivations that follow.}.
The function $H_{P}(z)$ is the $z$-transfer function of the
polyphase section decimating by $D_1$, whereas $H_{N}(z)$ is the
$z$-transfer function of the non recursive filter decimating by
$D_2$. The latter function is defined as
%
\begin{eqnarray}
\label{HN_z_def}
H_{N}(z) &=&\prod_{i=p_p+1}^{p-1}\left[ 1+r_i\cdot\left( z^{-2^i}+
z^{-2\cdot 2^i}\right)+z^{-3\cdot 2^{i}} \right]
\end{eqnarray}
whereby the coefficients $r_i$ are defined as
\figuramedia{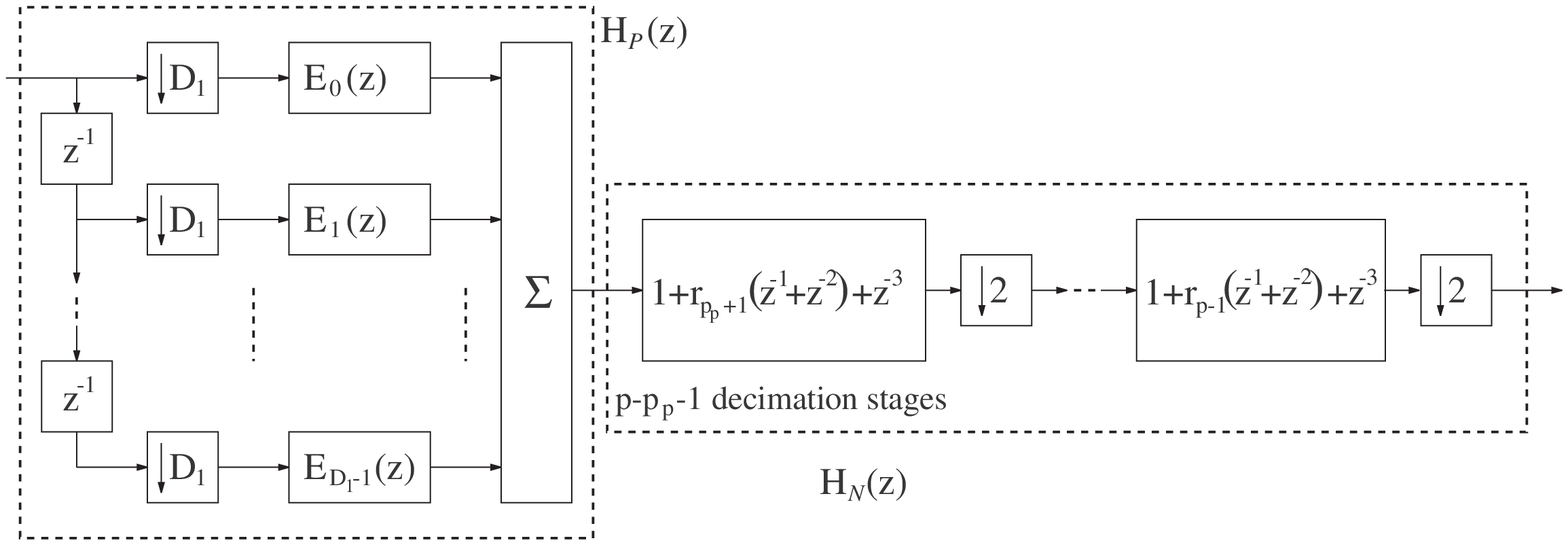}{Architecture of the
partial polyphase implementation of the decimation filter
$H(z)$.}{n_rec_impl_gcf3}
\begin{equation}\label{moltiplicatori_ru}
r_i=1+2\cos\left(2^i\alpha\right), ~\forall i=p_p+1,\ldots,p-1,
\end{equation}
whereas $\alpha=2\cdot 0.79\cdot \pi f_c^{i-1}$. We notice in
passing that according to the commutative property in \cite{Chu},
the filter $H_{N}(z)$ can be realized with the cascade of
$p-p_p-1$ stages, each one decimating by $2$. The integer $p_p$
can take on any value in the set $\left\{-1,\ldots,p-1\right\}$.

After some algebra, the frequency response of the filter $H_N(z)$
can be evaluated by substituting $z=e^{j\omega}$
in~(\ref{HN_z_def}):
\begin{equation}\label{third_order_gcf_fr}
\begin{array}{ll}
H_{N}(e^{j\omega}) = 2 \prod_{i=p_p+1}^{p-1}e^{-j3\cdot
2^{i-1}\omega}\cdot&\\
\cdot\left[ \cos\left(3\cdot 2^{i-1}\omega\right) +r_i \cos\left(
2^{i-1}\omega\right) \right],
\end{array}
\end{equation}
whereby $\omega=2\pi f_d$, and $r_i$ is defined
in~(\ref{moltiplicatori_ru}).

The impulse response $h_{P}(n),\forall n\in [0,3D_1-3],$ whose
$z$-transfer function is identified by $H_{P}(z)$, is defined as
\cite{laddomada_part_poly}:
\begin{eqnarray}\label{inv_z_trans_6_copia}
h_{P}(n)=e^{+j\alpha n}\sum_{k_3=0}^{n}e^{-2j\alpha
k_3}\sum_{k_2=0}^{k_3}e^{j\alpha k_2} \sum_{k_1=0}^{k_2}x_t(k_1).&
\end{eqnarray}
The definition of the sequence $x_t(n)$
in~(\ref{inv_z_trans_6_copia}) is
\begin{equation}\label{inv_z_trans_X_tz_copia}
x_t(n)= \delta(n)-r\delta(n-D_1)+r\delta(n-2D_1)-\delta(n-3D_1),
\end{equation}
whereby $r=1+\cos (\alpha D_1)$ and $\alpha=q 2\pi f_c^{i-1}$.

The impulse response in~(\ref{inv_z_trans_6_copia}) is used to
obtain the polyphase components
\begin{equation}\label{polifase_generale_2}
e_k(n)=h_{P}(D_1 n+k),~\forall k\in [0,D_1-1]
\end{equation}
of the filters $E_k(z)$ in the architecture shown in
Fig.~\ref{n_rec_impl_gcf3}.

Let us spend a few words about the parameters noticed in $H(z)$.
The choice $p_p=p-1$ allows the GCF filter $H(z)$ to be fully
realized in polyphase form, whereas the value $p_p=-1$ is such
that the filter is realized as the cascade of $p$ non recursive
decimation stages, each one decimating by $2$. Any intermediate
value of $p_p\in \{0,\ldots,p-2\}$ yields the partial polyphase
decomposition depicted in Fig.~\ref{n_rec_impl_gcf3}.

\noindent The first polyphase decimation stage allows the
reduction of the sampling rate by $D_1$, thus reducing the
operating rate of the subsequent decimation stages belonging to
$H_N(z)$. Any stage of $H_N(z)$ in Fig.~\ref{n_rec_impl_gcf3} is
constituted by a simple FIR filter operating at a different data
rate. Such an example, the $i$-th stage, with $i\in
\{0,\ldots,p-p_p-2\}$, is characterized by the transfer function
$\left[1+r_i\left(z^{-1}+z^{-2}\right)+z^{-3}\right]$ operating at
rate $f_s/\left(D_1\cdot 2^{i}\right)$, where $f_s$ is the
sampling frequency of the A/D converter.
\section{Design of Fixed-Point GCF Filters}
\label{Sensitivity_Analysis_section}
In this section we consider the problem of evaluating the
sensitivity of the filter $H(z)$ in~(\ref{third_order_gcf3_2_2})
with respect to the coefficients enclosed in both $H_P(z)$ and
$H_N(z)$. Then, the sensitivity function is employed in a design
algorithm that defines statistically the size of the registers in
the fixed-point implementation of the filter in such a way that
the error function in the frequency domain between the filter
$H(e^{j\omega})$ and the filter $H_q(e^{j\omega})$, which employs
quantized coefficients, is within given bounds with a preassigned
probability.

We show that the proposed framework gives a precise answer on the
choice of the proper split of the decimation factor $D$ between
the two substages $H_P(z)$ and $H_N(z)$. It is anticipated that
the best solution from a sensitivity point of view consists in
implementing the filter $H(z)$ without the polyphase stage
$H_P(z)$, i.e., with $D=D_2$ and $D_1=1$.

Before proceeding further, let us derive some observations on the
sensitivity function employed throughout this section. Given a
frequency response $H\left(e^{j\omega}\right)$, the sensitivity
analysis is usually accomplished on the magnitude of
$H\left(e^{j\omega}\right)$ with respect to its coefficients.
However, the sensitivity analysis based on the use of the
frequency response $H\left(e^{j\omega}\right)$ can be derived much
easier than the one that employs the function
$\left|H\left(e^{j\omega}\right)\right|$ \cite{Diniz_book}.
Moreover, the sensitivity function related to
$H\left(e^{j\omega}\right)$ provides an upper bound to the one
related to $\left|H\left(e^{j\omega}\right)\right|$. Consider the
frequency response
\[
H\left(e^{j\omega}\right)=\left|H\left(e^{j\omega}\right)\right|e^{j\varphi\left(H\left(e^{j\omega}\right)\right)},
\]
and a tagged multiplier $m$ belonging to
$H\left(e^{j\omega}\right)$. Then, the derivative of
$H\left(e^{j\omega}\right)$ with respect to $m$ can be evaluated
as follows:
\[
\begin{array}{lll}
\frac{\partial H\left(e^{j\omega}\right)}{\partial
m}&=&\frac{\partial
\left|H\left(e^{j\omega}\right)\right|}{\partial
m}e^{j\varphi\left(H\left(e^{j\omega}\right)\right)}+\\
&&+j\left|H\left(e^{j\omega}\right)\right|e^{j\varphi\left(H\left(e^{j\omega}\right)\right)}
\frac{\partial
\varphi\left(H\left(e^{j\omega}\right)\right)}{\partial m}.
\end{array}
\]
Upon observing that both $\left|H\left(e^{j\omega}\right)\right|$
and $\varphi\left(H\left(e^{j\omega}\right)\right)$ are real
functions of $\omega$, the bound
\begin{eqnarray}\small
\left|\frac{\partial H\left(e^{j\omega}\right)}{\partial
m}\right|&=&\sqrt{\left(\frac{\partial \left|H\right|}{\partial
m}\right)^2+\left|H\right|^2\left(\frac{\partial
\varphi\left(H\right)}{\partial m}\right)^2}\nonumber\\
&\ge&\left|\frac{\partial \left|H\right|}{\partial m}\right|
\end{eqnarray}
easily follows. Owing to this result, the sensitivity function
used in this work relies on the derivatives of the frequency
response $H\left(e^{j\omega}\right)$ with respect to its
multipliers.

When the multipliers belonging to $H\left(e^{j\omega}\right)$ are
quantized by employing rounding, the magnitude of the frequency
response becomes:
\begin{equation}\label{hgcf3_diff_2}
\left|H_q(e^{j\omega})\right|=\left|H(e^{j\omega})\right|+\Delta
\left|H(e^{j\omega})\right|,
\end{equation}
whereby $\Delta \left|H(e^{j\omega})\right|$ is an error function
that measures the distortion of the ideal frequency response
$\left|H(e^{j\omega})\right|$ from the one employing quantized
coefficients. Recalling the definition of the folding bands given
in~(\ref{folding_bands_def}), a key observation in the proposed
framework is that the error function $\Delta
\left|H(e^{j\omega})\right|$ must be properly bounded only within
the folding bands, concisely identified by $\textrm{FB}$.
Therefore, care must be devoted to the sensitivity analysis only
within the folding bands, while the behaviour of the error
function outside the FB does not affect the proposed fixed-point
design. Notice that these considerations only hold for the design
of decimation filters in multistage architectures, and cannot be
extended to the design of classical FIR filter.

The aforementioned considerations can be formalized as follow:
\begin{equation}\label{hgcf3_diff_2_1}
\left|\Delta \left|H(e^{j\omega})\right|\right|\le \chi (\omega),
\forall \omega\in \textrm{FB},
\end{equation}
whereby $\chi (\omega)$ is a suitable--positively defined--
tolerance function. Even though we can theoretically choose to
differentiate the behaviour of the tolerance function $\chi
(\omega)$ among the various folding bands, we do not pursue this
approach in this work. Therefore, the functions $\chi (\omega)$
used in the following will be constant functions across the
folding bands.

Next line of pursuit consists in investigating a statistical
technique in order to identify the word-lengths of the filter
coefficients. To this end, assume that the frequency response
$H(e^{j\omega})$ contains $N$ coefficients rounded by employing
the same fixed-point resolution ($b$-bit rounding quantization).
After quantization, each multiplier $m_i$ can be written as
$$m_{q,i}=m_i+\Delta m_i.$$ Therefore, the quantization of the $N$
coefficients yields $N$ zero-mean, statistically independent
random variables $\Delta m_i$ that are identically and uniformly
distributed in $\left[-2^{-2b}/2,+2^{-2b}/2\right]$
\cite{CrochiereOppenheim}. Under these hypotheses, and remembering
the relation (\ref{hgcf3_diff_2}), the variance of the error
function $\Delta\left| H(e^{j\omega})\right|$ can be evaluated as
follows:
\begin{equation}\label{variance_deltaH}
\sigma^2_{\Delta\left| H(e^{j\omega})\right|}\approx
\sigma^2_{\Delta m}\sum_{i=1}^{N}\left|\frac{\partial
H\left(e^{j\omega}\right)}{\partial m_i}\right|^2=\sigma^2_{\Delta
m}S_T(e^{j\omega}),
\end{equation}
whereby $\sigma^2_{\Delta m}=2^{-2b}/12$ is the variance of the
random variable $\Delta m$ under the hypothesis to employ rounding
to the nearest quantization level.

Owing to the condition $N\gg 1$, the error function $\Delta\left|
H(e^{j\omega})\right|$ can be modeled as a zero-mean Gaussian
random variable with variance given by (\ref{variance_deltaH})
\cite{Crochiere_StatApproach}-\cite{Avenhaus}. Therefore, we can
estimate the probability $$p=P\left[\left|\Delta\left|
H\right|\right|\le y\sigma_{\Delta\left|
H(e^{j\omega})\right|}\right]$$ that $\Delta\left|
H(e^{j\omega})\right|$ falls within a proper interval, say from
$-y\sigma_{\Delta\left| H(e^{j\omega})\right|}$ to
$+y\sigma_{\Delta\left| H(e^{j\omega})\right|}$, as follows:
\begin{equation}\small\label{prob_deltaH}
\frac{1}{\sqrt{2\pi}\sigma_{\Delta\left|
H(e^{j\omega})\right|}}\int_{-y\sigma_{\Delta\left|
H(e^{j\omega})\right|}}^{+y\sigma_{\Delta\left|
H(e^{j\omega})\right|}}e^{-\frac{x^2}{2\sigma^2_{\Delta\left|
H(e^{j\omega})\right|}}}dx.
\end{equation}
It is convenient to employ the new variable
$$z=\frac{x}{\sqrt{2}\sigma_{\Delta\left| H(e^{j\omega})\right|}}$$ in~(\ref{prob_deltaH}), thus obtaining
\begin{equation}\label{prob_deltaH_subst}
p=P\left[\left|\Delta\left| H\right|\right|\le
y\sigma_{\Delta\left|
H(e^{j\omega})\right|}\right]=\frac{2}{\sqrt{\pi}}\int_{0}^{\frac{y}{\sqrt{2}}}e^{-z^2}dz.
%
%
\end{equation}
Let us spend few words about the result (\ref{prob_deltaH_subst}).
The term $p$ is the probability that the magnitude of the error
function $\left|\Delta \left|H(e^{j\omega})\right|\right|$
in~(\ref{hgcf3_diff_2_1}) is upper-bounded by
$y\sigma_{\Delta\left| H(e^{j\omega})\right|}$.

The relation between $p$ and $y$ in~(\ref{prob_deltaH_subst}) is
illustrated in Fig.~\ref{prob}. As an instance, the value $y=2$
has to be chosen in order to guarantee with a probability equal to
$95\%$ that the error function is bounded by
$2\sigma_{\Delta\left| H(e^{j\omega})\right|}$ in the frequency
domain.

How can we employ this result in a practical design? Upon
recalling (\ref{hgcf3_diff_2_1}), and given a proper $p$, we
choose $y$ in such a way that the following relation holds:
\figuragrossa{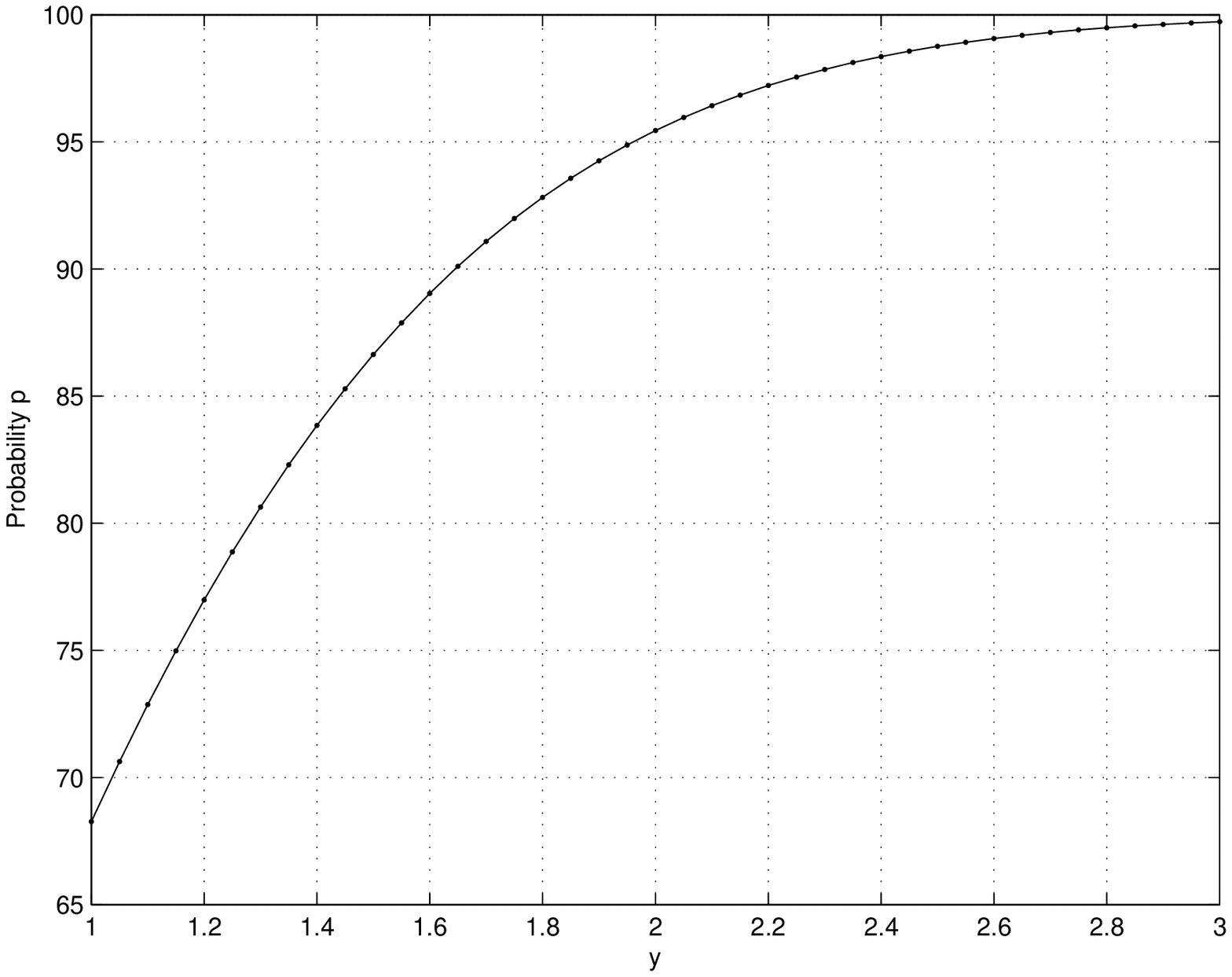}{Relation between the probability $p$ and
the parameter $y$ in~(\ref{prob_deltaH_subst}).}{prob}
\[
y\cdot\sigma_{\Delta\left| H(e^{j\omega})\right|}\approx \chi
(\omega)\Rightarrow y\cdot\sigma_{\Delta
m}\sqrt{S_T(e^{j\omega})}\approx \chi (\omega).
\]
This relation can be rewritten as
\begin{equation}\label{prob_deltaH_1}
%
y \frac{2^{-b}}{\sqrt{12}}\sqrt{S_T(e^{j\omega})}\approx \chi
(\omega).
\end{equation}
By doing so, we guarantee that (\ref{hgcf3_diff_2_1}) is verified
with probability $p$ given by (\ref{prob_deltaH_subst}).

Given this statistical framework, let us focus on the fixed-point
design of the considered GCF filter.

Assume that the filter coefficients are represented with the
following fixed-point notation: $I_n$ bits are devoted to the
integer part of the coefficients, while $F_n$ is the number of
bits devoted to the fractional part. Therefore, the size of the
filter coefficients is equal to $1+I_n+F_n$ bits, accounting for
the sign of the number.

The next two subsections derive the sizes of both $I_n$ and $F_n$
in the fixed-point implementation.
\subsection{Evaluation of the fractional size, $F_n$}
\noindent Considering $b=F_n$, and solving (\ref{prob_deltaH_1})
for $F_n$, we can obtain the size $F_n$ of the fractional part
%
%
%
in order for (\ref{hgcf3_diff_2_1}) to hold with a probability $p$
given by (\ref{prob_deltaH_subst}):
%
%
%
\figuramedia{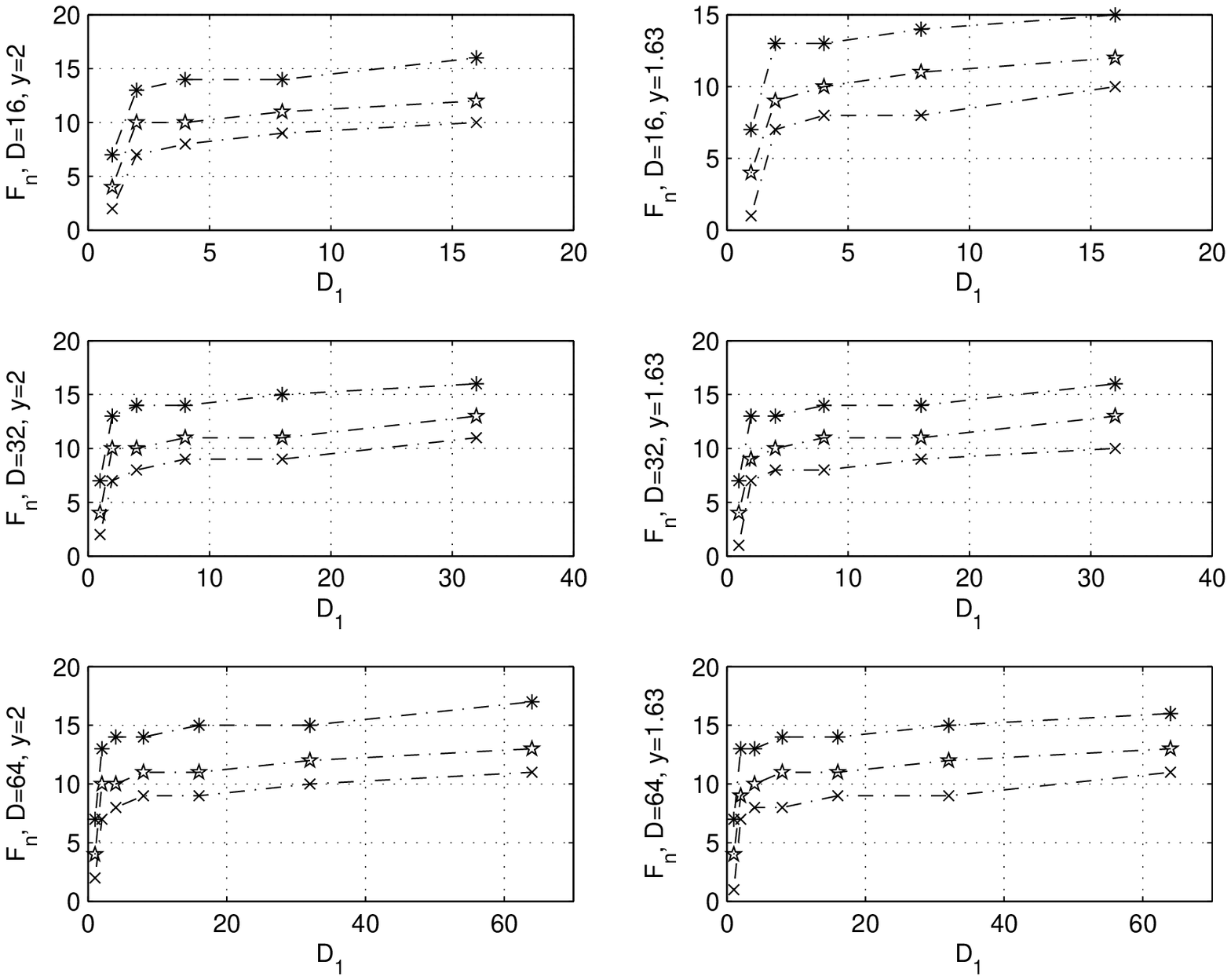}{Behaviour of $F_n$ in~(\ref{size_F_n}) as a
function of the decimation factor $D_1$ of the polyphase stage for
three values of the function $\chi (\omega)$, assumed constant
across the folding bands. Left subplots are associated to the
value $y=2$ in~(\ref{prob_deltaH_subst}) corresponding to the
probability $p=95\%$, whereas the rightmost subplots are
associated to the value $y=1.63$, which corresponds to the
probability $p=90\%$. The three curves in each subplot are
parameterized with respect to the value of the function $\chi
(\omega)$ as follows: curves labeled with the mark $-\times-$ are
for $\chi (\omega)=5\times 10^{-3}$, the mark $-\star-$ is
associated to $\chi (\omega)=10^{-3}$, and mark $-*-$ is used for
$\chi (\omega)=10^{-4}$.}{Fn}
\begin{equation}\label{size_F_n}
F_n=\left\lceil -\log_2\left[\sqrt{12}\min_{\omega\in
\textrm{FB}}\frac{\chi
(\omega)}{y\sqrt{S_T(e^{j\omega})}}\right]\right\rceil,
\end{equation}
whereby the minimum is taken only over the folding bands derived
in~(\ref{folding_bands_def}), and $\left\lceil\cdot\right\rceil$
is the ceil of the underlined number.

The evaluation of the fractional part $F_n$ in~(\ref{size_F_n})
relies on the sensitivity function $S_T\left(e^{j\omega}\right)$.
To keep the presentation concise, the derivation of the
sensitivity function is reported in the Appendix.

The behaviour of $F_n$ in~(\ref{size_F_n}) as a function of the
decimation factor $D_1$ of the polyphase stage is illustrated in
Fig.~\ref{Fn} for various values of the function $\chi (\omega)$.
The setup for deriving the results in Fig.~\ref{Fn} is as follows.
We considered a two-stage decimating architecture ($m=2$ in
Fig.~\ref{arch}) and an oversampling ratio $\rho=4\cdot D$, where
$D$ is the other parameter associated to each subplot in
Fig.~\ref{Fn}. The first decimation filter in the 2-stage
architecture is the investigated GCF filter, thus
$H_i(z)\left|_{i=1}\right.=H(z)$ in Fig.~\ref{arch}.

The normalized digital bandwidth of the useful signal at the input
of the GCF decimation filter is
$f_c^{i-1}\left|_{i=1}\right.=f_c^o=\frac{1}{2\rho}$. This is also
half the width of the folding bands seen by the first decimation
filter in the two-stage architecture. Therefore, the value of
$\alpha$ appearing in the definition of filters $H_P(z)$ and
$H_N(z)$ in~(\ref{HN_z_def}) and (\ref{inv_z_trans_6_copia}) is
$0.79\cdot 2\pi f_c^{o}=0.79\pi/\rho$. As a note aside, notice
that the value of $\alpha$ would be different if the GCF filter
were used in the second stage of the multistage chain in
Fig.~\ref{arch}, due to the different value of $f_c^{i-1}$.

We considered three different constant functions\footnote{We
notice in passing that other behaviours can be associated to the
tolerance function $\chi (\omega)$, depending on the required
sensitivity desired in the various folding bands. However, we do
not pursue this approach in this work, and assume that all the
folding bands affect equally the size $F_n$ of the fractional
part.} $\chi (\omega)$ in order to draw (\ref{size_F_n}), namely
$\chi (\omega)=5\times 10^{-3}$, $\chi (\omega)=10^{-3}$, and
$\chi (\omega)=10^{-4}$.

Each subplot is associated to a specific decimation factor
$D=D_1\cdot D_2$. Therefore, given the constant $D$ noticed in the
ordinate of each subplot, the number of stages belonging to
$H_N(z)$ is reduced as long as $D_1$ increases. In particular, the
abscissa $D_1=1$ in each subplot is associated to the case
$D=D_2$, which means that the GCF filter is implemented without
the polyphase stage.

Moreover, we consider two different values of $y$ related to the
probabilities $p=90\%$ and $p=95\%$ illustrated in
Fig.~\ref{prob}.

Some observations are in order.
\begin{itemize}
    \item A comparison among the leftmost and the rightmost
    subplots in Fig.~\ref{Fn} reveals the need of
    one additional bit for the fractional part in order to guarantee that the constraint on the
    tolerance function is attained in the frequency domain with
    probability $95\%$ with respect to the case $90\%$.
    \item For given $y$, $D$, and $D_1$ in the abscissa, the number $F_n$ of fractional
    bits increases as long as a lower tolerance function
    $\chi(\omega)$ is desired.
    \item Given $D$, the size $F_n$ of the fractional part of the fixed-point
    implementation increases as long as the number of cascaded
    cells in $H_N(z)$ decreases. This is equivalent to say that $F_n$ increases as long as
    $D_1$ does. This observation suggests that the GCF filter
    $H(z)$ implemented as $H_N(z)$, i.e., without the polyphase
    stage, allows to contain the computational complexity of the
    GCF filter.
\end{itemize}
\figuramedia{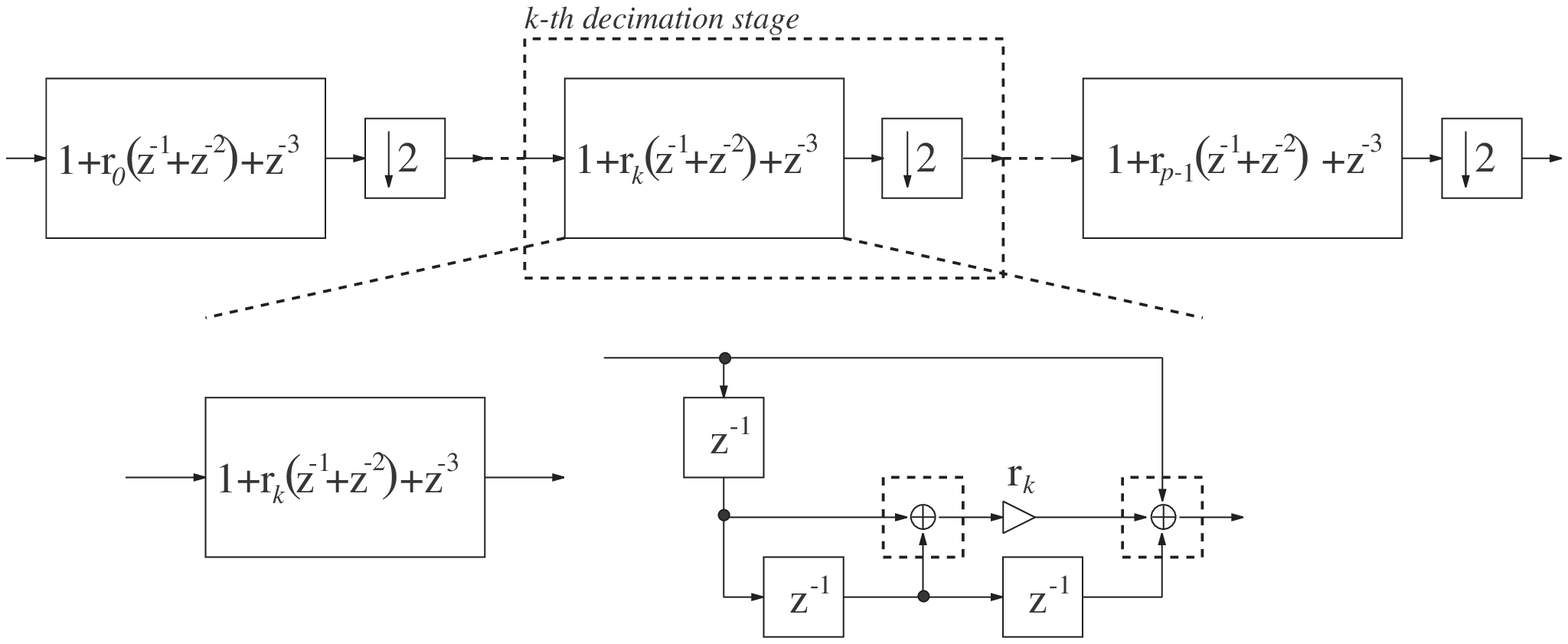}{Architecture of a
non-recursive implementation of the decimation filter $H(z)$.
Index $k$, which identifies the decimation stage, belongs to the
range $0,1,\ldots,\log_2 D-1=p-1$. The figure also shows an
effective implementation of the $k$-th filter with transfer
function $1+r_k(z^{-1}+z^{-2})+z^{-3}$.}{cascaded_architecture}
The latter observation above suggests that an effective
implementation of the GCF filter is $H(z)=H_N(z)$. Therefore, the
3rd-order GCF filter is realized with the cascaded architecture
shown in Fig.~\ref{cascaded_architecture}.

This architecture follows from (\ref{HN_z_def}) upon setting
$p_p=-1$:
\begin{eqnarray}
\label{HN_z_def_finale}
H_{N}(z) &=&\prod_{k=0}^{p-1}\left[ 1+r_k\cdot\left( z^{-2^k}+
z^{-2\cdot 2^k}\right)+z^{-3\cdot 2^{k}} \right]
\end{eqnarray}
whereby the coefficients $r_k$ are defined as
\begin{eqnarray}\label{moltiplicatori_ru_finale}
r_k&=&1+2\cos\left(2^k\alpha\right), ~\forall
k=0,\ldots,p-1\nonumber\\
\alpha&=&2\cdot 0.79\cdot \pi f_c^{i-1}.
\end{eqnarray}
Applying the commutative property \cite{Chu}, the cascaded
implementation shown in Fig.~\ref{cascaded_architecture} easily
follows.
\subsection{Evaluation of the integer part, $I_n$}
\noindent This section is focused on the evaluation of the size of
the integer part $I_n$ in the fixed-point representation of the
filter coefficients. To this goal, consider the architecture in
Fig.~\ref{cascaded_architecture}, and focus on the $k$-th
decimation stage. Let $I_n^k$ be the size of the integer part in
the $k$-th decimation stage.

%
The impulse response associated to the transfer function
$1+r_k(z^{-1}+z^{-2})+z^{-3}$ is $$h_k(n)=\delta (n)+r_k \delta
(n-1)+r_k \delta (n-2)+\delta (n-3).$$
Upon relying on general considerations about dynamic range
overflow, it is simple to observe that the worst-case dynamic
range growth $G_k$ of the $k$-th stage is
\[
G_k\le
\log_2\left(\sum_{n=0}^{3}|h_k(n)|\right)=\log_2\left(2+2r_k\right)\le
3,
\]
where the last inequality stems from the observation
$$r_k=1+2\cos\left(2^k\alpha\right)\le 3, \forall
k=0,\ldots,p-1.$$
Therefore, the size of the integer part $I^k_n$ (in bits) that
avoids overflow, is equal to the sum between the width of the
input word (in bits) and $G_k$.
\section{Simulation Results}
\label{Comparisons_Simulation_Results_Section}
In this section, we compare some GCF filters designed with the
framework proposed in the previous section, with classical comb
filters. We also provide a set of simulation results obtained by
employing a fixed-point realization of a 3rd-order GCF filter for
decimating a discrete-time signal oversampled by a $\Sigma\Delta$
A/D converter.

The first set of results is proposed to compare the frequency
response of a 3rd-order GCF filter $H(e^{j\omega})$ with the one
obtained with a fixed-point implementation. Let us summarize the
setup. We consider a two-stage decimating architecture ($m=2$ in
Fig.~\ref{arch}), whereby the first stage employs a GCF filter
decimating by $D=D_2=16$ (i.e., the GCF filter is implemented with
the architecture shown in Fig.~\ref{cascaded_architecture} with
$p=4$), whereas the second stage presents a decimation factor
equal to $4$. With this setup, the oversampling ratio is
$\rho=64$.

From the upper-leftmost subplot in Fig.~\ref{Fn}, we notice that
$F_n=7$ in order to satisfy the bound $$\left|\Delta
\left|H(e^{j\omega})\right|\right|\le \chi (\omega)=10^{-4},
\forall \omega\in \textrm{FB},$$ with a probability equal to
$95\%$.

The normalized digital bandwidth of the useful signal at the input
of the GCF decimation filter is
$$f_c^{i-1}\left|_{i=1}\right.=f_c^o=\frac{1}{2\rho}=\frac{1}{128}.$$
The magnitude of the frequency response (dotted-line curve) of the
3rd-order filter $H(e^{j\omega})$ without coefficients'
quantization is shown in Fig.~\ref{Comparison_Freq_Resp} for
$D=16$. From (\ref{folding_bands_def}), we notice the presence of
the following folding bands:
\figuramatlab{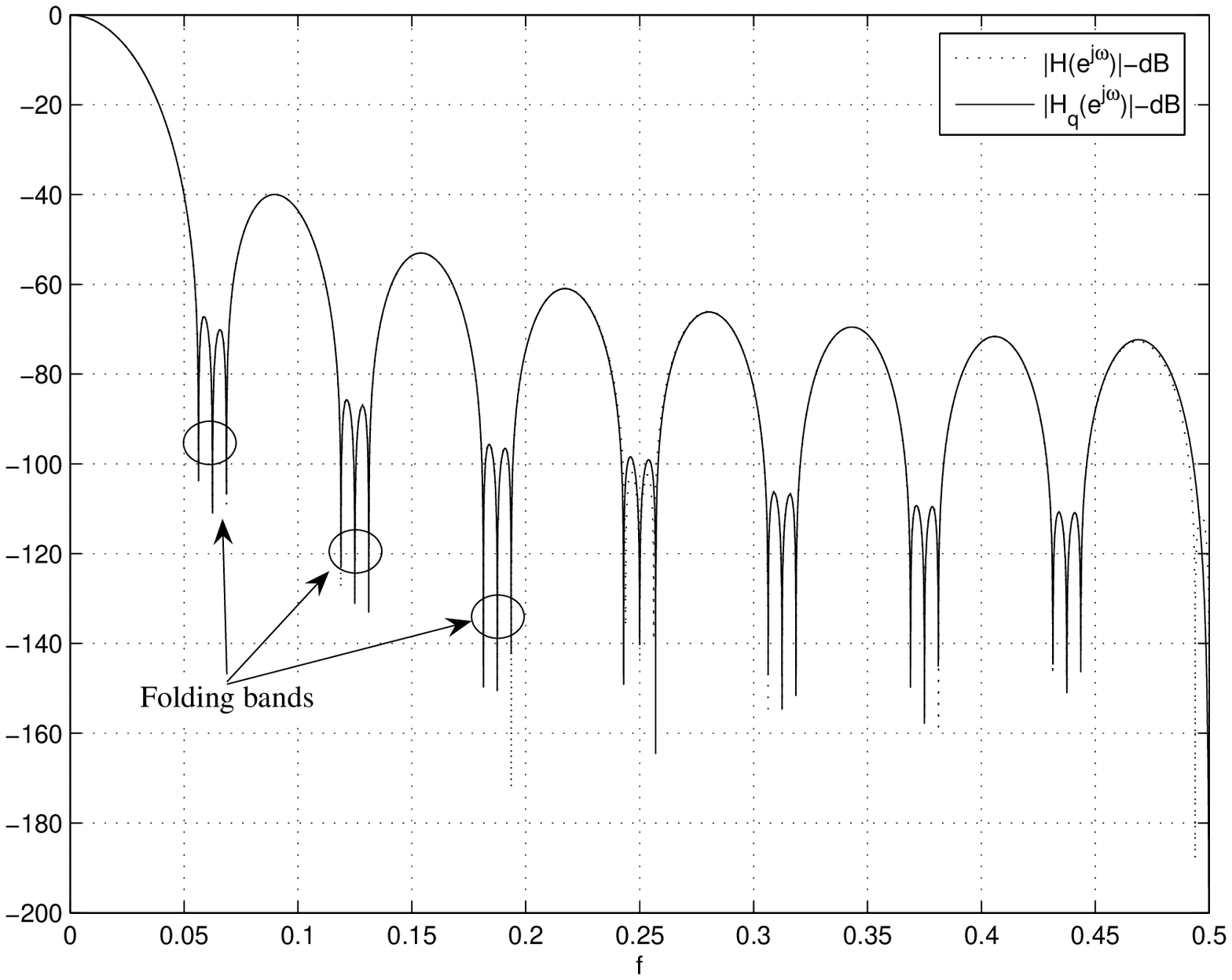}{Magnitude of the frequency
response of filter $H(e^{j\omega})$ employing real coefficients
(continuous curve), and frequency response of filter
$H_q(e^{j\omega})$ employing fixed-point coefficients (dotted
curve). Both frequency responses are mostly superimposed. The
decimation factor is $D=16$.}{Comparison_Freq_Resp}
\[
\begin{array}{lll}
\left[\frac{k}{16}-\frac{1}{128};\frac{k}{16}+\frac{1}{128}\right],&k=1,\ldots,k_M=8,&
\end{array}
\]
\noindent some of which have been highlighted in
Fig.~\ref{Comparison_Freq_Resp}. In the same figure, we show for
comparison the frequency response (continuous curve) of the GCF
filter whereby the coefficients have been quantized with $F_n=7$,
as discussed above. Notice that the two frequency responses are
mostly superimposed, thus confirming the effectiveness of the
proposed design framework.

The frequency response $H_q(e^{j\omega})$ is compared with the one
of a classical 3rd-order comb filter in
Fig.~\ref{Comparison_Freq_Resp_comb}. The figure clearly
highlights the behavior of the GCF filter across the folding
bands: unless a classical 3rd-order comb that places a 3rd-order
zero in the frequencies $\frac{k}{16},~\forall k=1,\ldots,k_M=8,$
the GCF filter places its zeros in the frequencies
$$\frac{k}{16};~\frac{k}{16}-0.79\cdot
f_c^o;~\frac{k}{16}+0.79\cdot f_c^o;~\forall k=1,\ldots,k_M=8.$$
The last set of results is obtained by resorting to simulation.
Employing Matlab, we simulated a $2$nd order $\Sigma\Delta$
converter with a 2-level quantizer and a sampling frequency
$f_s=25.6$~kHz. The input signal is a band-limited random signal
with bandwidth $f_x=100$~Hz. From the values of $f_s$ and $f_x$,
it is $\rho=128$, while the normalized digital bandwidth of the
sampled signal is $f^o_c=\frac{1}{2\rho}=\frac{1}{256}$.

The oversampled signal is then decimated by $D=16$ employing a
3rd-order GCF filter\footnote{We recall that the order of the GCF
filter has to be greater or equal to $B+1$, whereby $B\ge 1$ is
the order of the $\Sigma \Delta$ modulator
\cite{laddomada_sharp}.}. The power spectrum of the digital signal
at the output of the $\Sigma\Delta$ A/D converter is shown in the
upper subplot of Fig.~\ref{simulazione_1}. Notice that, as
expected, the useful signal with bandwidth $f^o_c=\frac{1}{256}$
is shrunk at baseband, while the $\Sigma\Delta$ A/D converter has
pushed the noise power spectrum outside the useful signal
bandwidth $[0,f^o_c]$. The power spectrum of the decimated signal
is shown in the lower subplot of Fig.~\ref{simulazione_1}. Notice
that the useful signal bandwidth is now $f_c^{1}=f_c^o\cdot
D=\frac{16}{256}\approx 0.063$.
%
%
%
\figuramatlab{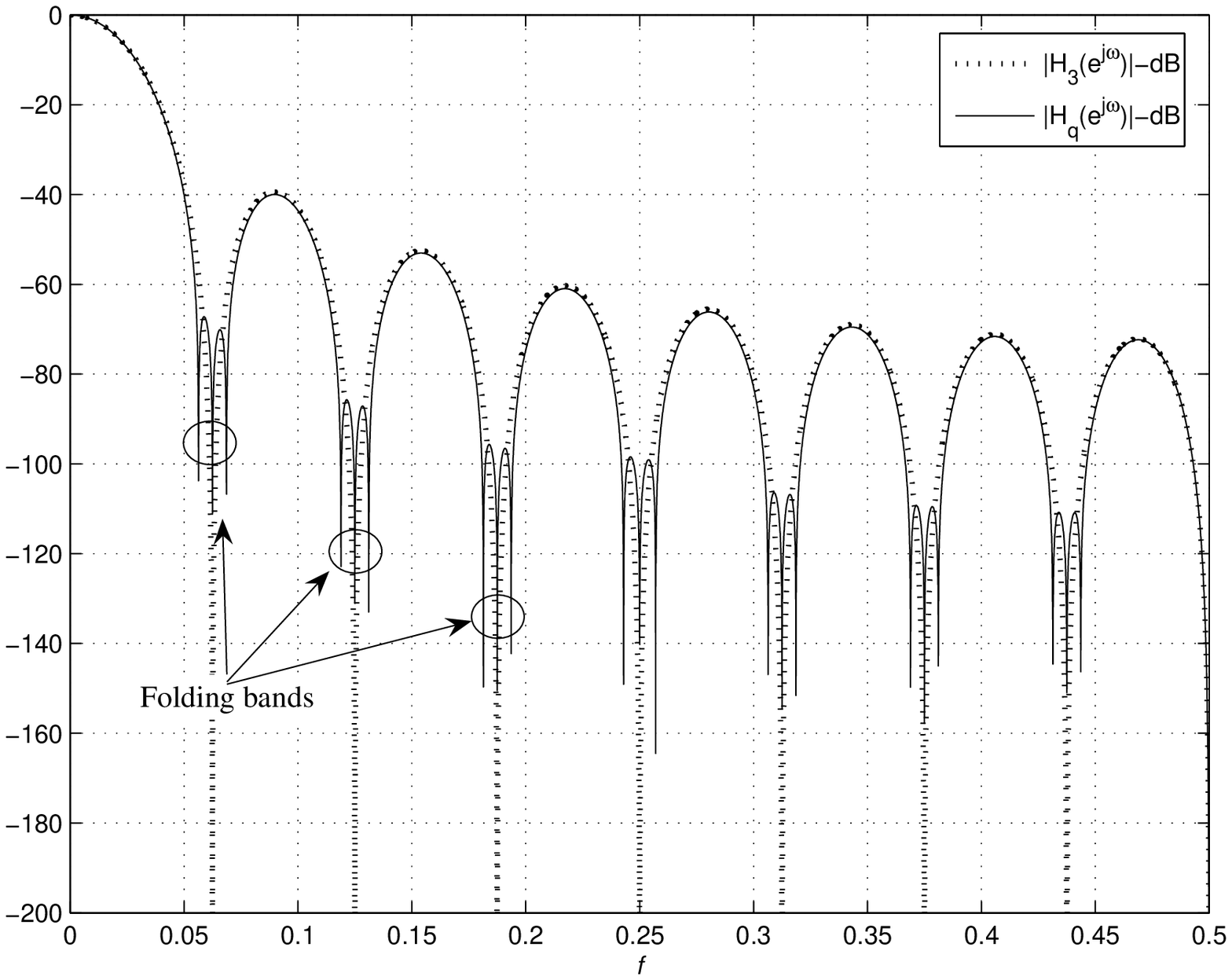}{Magnitude of the
frequency response of a classical $3$-rd-order comb filter
$H_3(e^{j\omega})$ (dotted curve), and frequency response of
filter $H_q(e^{j\omega})$ employing fixed-point coefficients
(continuous curve). The decimation factor is
$D=16$.}{Comparison_Freq_Resp_comb}
\section{Conclusions}
\label{conclusions}
This paper focused on the design of computationally efficient
Generalized Comb Filters (GCF), i.e., anti-aliasing filters that,
employed as decimation filters in multistage architectures,
guarantee superior performance in terms of selectivity and
quantization noise rejection compared to classical comb filters.
GCF filters can be realized by relying on both IIR and FIR
architectures, even though FIR schemes do not present instability
problems stemming from coefficients' quantization.

As a reference filter in the class of GCF filters, a third order
FIR architecture, realized by employing a partial polyphase
architecture, was used throughout the paper. We proposed a
sensitivity analysis in order to first investigate the effects of
the coefficients' quantization on the frequency response of the
designed filters, and, then, to define the registers'lengths in
the proposed fixed-point implementation.
%
%
%
\figuramatlab{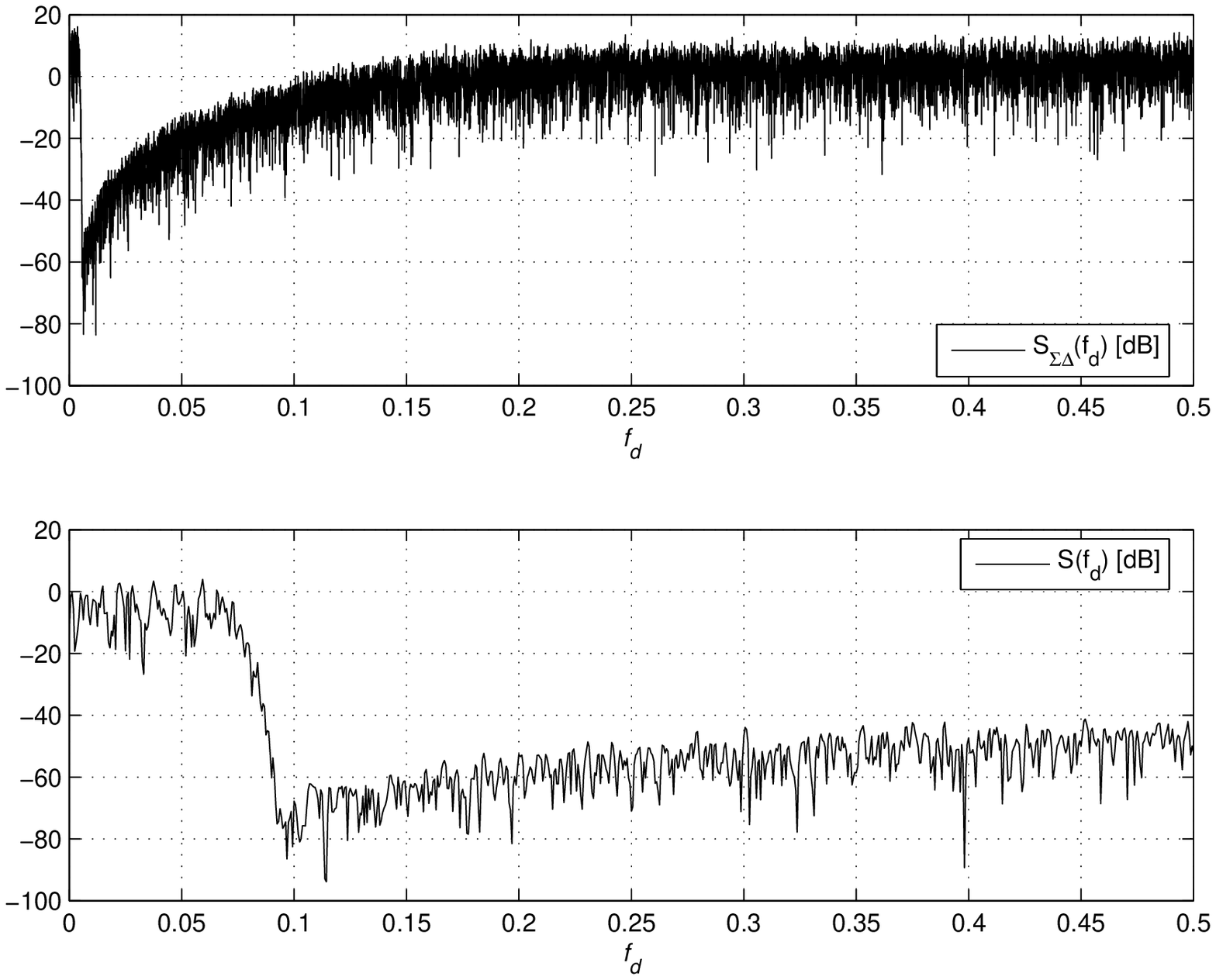}{Power spectrum of the digital
signal at the output of a 2nd-order $\Sigma\Delta$ A/D converter
sampling at $f_s=25.6$~kHz (upper subplot), and power spectrum of
the signal decimated by $D=16$ with a 3rd-order GCF filter (lower
subplot).}{simulazione_1}
%
%

%
\section*{Appendix}
In this Appendix we derive the sensitivity function
$S_T\left(e^{j\omega}\right)$ needed in the evaluation of the
fractional part $F_n$ in~(\ref{size_F_n}). Let us evaluate the
function $S_T\left(e^{j\omega}\right)$ in~(\ref{variance_deltaH})
for the considered partial polyphase architecture. To this end, we
consider three different cases
depending on the value of $p_p$.\\

\noindent \textbf{First Case: $p_p=-1$.} This is the case in which
$D_1=1$ and $D=D_2$. The frequency response of the GCF filter is
$H\left(e^{j\omega}\right)=H_N\left(e^{j\omega}\right)$, and the
filter is implemented without the polyphase stage. Therefore, the
sensitivity function $S_T\left(e^{j\omega}\right)$ can be
evaluated as follows:
\begin{eqnarray}
S_T\left(e^{j\omega}\right)&=&
\left|H_N\left(e^{j\omega}\right)\right|^2\sum_{i=0}^{p-1}\left|\frac{\cos\left(3\cdot
2^{i-1}\omega\right)}{\cos\left(2^{i-1}\omega\right)}+r_{i}
 \right|^{-2}.
\end{eqnarray}
The previous equation stems from (\ref{third_order_gcf_fr}) upon
noting that, for $p_p=-1$, the derivative of $H_{N}(e^{j\omega})$
with respect to $r_u,~\forall u=0,\ldots,p-1$, can be evaluated
as:
\begin{equation}\label{sensitivity_hgcf_3_2}
\begin{array}{ll}
\frac{\partial H_{N}(e^{j\omega}) }{\partial r_u}=2e^{-j3\cdot
2^{u-1}\omega}\cos\left(
2^{u-1}\omega\right)\cdot&\\
\prod_{m=0,~m\ne u}^{p-1}e^{-j3\cdot 2^{m-1}\omega} \left[
\cos\left(3\cdot 2^{m-1}\omega\right) +r_m\cdot \cos\left(
2^{m-1}\omega\right) \right].&\nonumber
\end{array}
\end{equation}
%
%
%
By multiplying and dividing for the function $$\cos\left(3\cdot
2^{u-1}\omega\right)+r_u\cdot \cos\left( 2^{u-1}\omega\right),$$
and
recalling~(\ref{third_order_gcf_fr}),~(\ref{sensitivity_hgcf_3_2})
can be rewritten as follows:
\begin{equation}\label{sensitivity_hgcf_3_2_2}
\begin{array}{ll}
\frac{\partial H_{N}(e^{j\omega}) }{\partial
r_u}=H_{N}(e^{j\omega})\cdot\frac{\cos\left(2^{u-1}\omega\right)
}{\cos\left(3\cdot 2^{u-1}\omega\right)+r_u\cdot \cos\left(
2^{u-1}\omega\right)}.&
\end{array}
\end{equation}
\\
\noindent \textbf{Second Case: $p_p=p-1$.} This is the case in
which $D_2=1$ and $D=D_1$. The frequency response of the GCF
filter is $H\left(e^{j\omega}\right)=H_P\left(e^{j\omega}\right)$,
and the filter is fully implemented with a polyphase architecture.

The polyphase decomposition of the $z$-transfer function
$H_{P}(z)$ is defined as follows:
\begin{equation}\label{third_order_gcf3_2_poly_dec}
\begin{array}{ll}
H_{P}(z) =\sum_{k=0}^{D_1-1}z^{-k}E_k\left(z^{D_1}\right).
\end{array}
\end{equation}
From (\ref{third_order_gcf3_2_poly_dec}) and
(\ref{polifase_generale_2}), $H_{P}(e^{j\omega})$ can be rewritten
as:
\begin{equation}\label{F_k_n}
H_{P}(e^{j\omega})=\sum_{k=0}^{D_1-1}\sum_{n=0}^{\left\lfloor
L/D_1\right\rfloor} h_P(D_1\cdot n+k) e^{-j\omega (D_1 n+ k)},
\end{equation}
which is valid $\forall n,k$ such that $0\le D_1\cdot n+k<L$.

Upon observing that
\[
\frac{\partial H_{P}(e^{j\omega}) }{\partial h_P}=e^{-j\omega (D_1
n+k)}, ~\forall n,k,
\]
the sensitivity $S_T\left(e^{j\omega}\right)$ reduces to:
\[
S_T\left(e^{j\omega}\right)=\sum_{i=1}^{L}\left|\frac{\partial
H_P}{\partial h_P(i)}\right|^2=L=3D_1-2.
\]
which corresponds to the number of multipliers in
$H_P(e^{j\omega})$.\\

\noindent \textbf{Intermediate Case: $p_p=0,\ldots,p-2$.} This is
the case where the GCF filter is implemented with the partial
polyphase architecture discussed above. Let $N=L+p-p_p-1$ be the
number of multipliers belonging to $H\left(e^{j\omega}\right)$
($L=3D_1-2$ is the number of multipliers belonging to
$H_P\left(e^{j\omega}\right)$, while $p-p_p-1$ is the number of
coefficients belonging to $H_N\left(e^{j\omega}\right)$).

The sensitivity function $S_T\left(e^{j\omega}\right)$ assumes the
following expression:
%
\begin{eqnarray}
\sum_{i=1}^{N}\left|\frac{\partial
H\left(e^{j\omega}\right)}{\partial
m_i}\right|^2&=&\sum_{i=1}^{L}\left|H_N\right|^2\left|\frac{\partial
H_P}{\partial m_i}\right|^2\\
& &+\sum_{i=1}^{p-p_p-1}\left|H_P\right|^2\left|\frac{\partial
H_N}{\partial m_i}\right|^2.\nonumber
\end{eqnarray}
%
After some algebra, the previous relation can be rewritten as:
\begin{eqnarray}\label{sens_funct_2}\small
S_T\left(e^{j\omega}\right)&=&L\cdot\left|H_N\left(e^{j\omega}\right)\right|^2+\left|H_P\left(e^{j\omega}\right)\right|^2\nonumber\\
&\cdot
&\left|H_N\left(e^{j\omega}\right)\right|^2\sum_{i=1}^{p-p_p-1}
\left|\frac{\cos\left(3\cdot
2^{p_p+i-1}\omega\right)}{\cos\left(2^{p_p+i-1}\omega\right)}+r_{p_p+i}\right|^{-2}\nonumber\\
&=&
L\left|H_N\left(e^{j\omega}\right)\right|^2+\left|H\left(e^{j\omega}\right)\right|^2\\
&&\sum_{i=1}^{p-p_p-1} \left|\frac{\cos\left(3\cdot
2^{p_p+i-1}\omega\right)}{\cos\left(2^{p_p+i-1}\omega\right)}+r_{p_p+i}\right|^{-2},\nonumber
\end{eqnarray}
whereby
\[
r_{p_p+i}=1+2\cdot \cos \left(2^{p_p+i}\alpha\right).
\]

\begin{thebibliography}{99}

\bibitem{Temes} S. R. Norsworthy, R. Schreier, and G. C. Temes,
{\em Delta-Sigma Data Converters, Theory, Design, and Simulation},
IEEE Press, 1997.

\bibitem{Hoge} E. B. Hogenauer, ``An economical class of digital filters for decimation and interpolation,'' {\em
IEEE Trans. on Ac., Speech and Sign. Proc.,} Vol. ASSP-29, pp.
155-162, No. 2, April 1981.

\bibitem{candy_decim} J.C. Candy, ``Decimation for sigma delta modulation,'' {\em
IEEE Trans. on Comm.}, Vol. COM-34, pp.72-76, No. 1, Jan. 1986.

\bibitem{Mitola} J. Mitola, ``The software radio architecture,'' {\em IEEE
Comm. Magazine,} Vol.33, No.5, pp. 26-38, May 1995.

\bibitem{Laddomada_PCbased} M. Laddomada, F. Daneshgaran, M. Mondin, and R.M. Hickling, ``A PC-based software receiver using a novel front-end technology,''
{\em IEEE Communications Magazine,}  Vol.39, No.8, pp.136-145,
Aug. 2001.

\bibitem{Laddomada_transceiver} F. Daneshgaran and M. Laddomada, ``Transceiver front-end technology for software radio implementation
of wideband satellite communication systems,'' {\em Wireless
Personal Communications, Kluwer,} Vol.24, No.12, pp. 99-121,
December 2002.

\bibitem{CrochiereRabiner} R. E. Crochiere and L. R. Rabiner, {\em Multirate Digital Signal Processing},
Prentice-Hall PTR, 1983.

\bibitem{dolecek_6} G. Jovanovic Dolecek (Editor), {\em Multirate systems: Design and
applications,} IGP, USA, 2001.

\bibitem{Chu} S. Chu and C. S. Burrus,
``Multirate filter designs using comb filters,'' {\em IEEE
Transactions on Circuits and Systems,} vol. CAS-31, pp. 913–924,
Nov. 1984.

\bibitem{laddomada_multistage_dec} M. Laddomada,
``Design of multistage decimation filters using cyclotomic
polynomials: Optimization and design issues,'' {\em To appear on
IEEE Trans. on Circuits and Systems I,} 2008.

\bibitem{CrochiereRabinerTut} R. E. Crochiere and L. R. Rabiner, ``Interpolation and decimation of digital signals--A tutorial review,'' {\em
Proceedings of the IEEE,} Vol.69, No.3, pp. 300-331, March 1981.

\bibitem{Vaidyanathan} P.P. Vaidyanathan, ``Multirate digital filters, filter banks, polyphase networks, and
applications: a tutorial,'' {\em Proceedings of the IEEE,} Vol.78,
No.1, pp. 56-93, Jan. 1990.

\bibitem{Coffey1} M.W. Coffey, ``Optimizing multistage decimation and interpolation processing-Part I,''
{\em IEEE Signal Proc. Letters,} Vol.10, No.4, pp. 107-110, April
2003.

\bibitem{Coffey2} M.W. Coffey, ``Optimizing multistage decimation and interpolation processing-Part II,''
{\em IEEE Signal Proc. Letters,} Vol.14, No.1, pp. 24-26, Jan.
2007.

\bibitem{Letizia} L. Lo Presti, ``Efficient modified-sinc filters for sigma-delta A/D converters,'' {\em
IEEE Trans. on Circ. and Syst.-II,} Vol. 47, pp. 1204-1213, No.
11, November 2000.

\bibitem{max} M. Laddomada, L. Lo Presti, M. Mondin, and
C. Ricchiuto, "An efficient decimation sinc--filter design for
software radio applications", {\em In Proc. of IEEE SPAWC}, March
20-23, 2001.
%

\bibitem{laddomada_gcf} M. Laddomada,
``Generalized comb decimation filters for $\Sigma \Delta$ A/D
converters: Analysis and design,'' {\em IEEE Trans. on Circuits
and Systems I,} Vol.54, No. 5, pp. 994-1005, May 2007.

\bibitem{gao} Y. Gao, J. Tenhunen, and H. Tenhunen, ``A fifth-order comb decimation filter for multi-standard
transceiver applications,'' {\em In Proc. of IEEE ISCAS 2000,} May
2000, pp. III-89-III-92.

\bibitem{aboushady} H. Aboushady, Y. Dumonteix, M. Lou$\ddot{e}$rat, and H. Mehrez,
``Efficient polyphase decomposition of comb decimation filters in
$\Sigma \Delta$ analog-to-digital converters,'' {\em IEEE Trans.
on Circ. and Syst.-II,} Vol. 48, pp. 898-903, No. 10, October
2001.


\bibitem{kwentus} A.Y. Kwentus, Z. Jiang, and A.N. Willson Jr.,
``Application of filter sharpening to cascaded integrator-comb
decimation filters,'' {\em IEEE Trans. on Signal Proc.,} Vol. 45,
pp. 457-467, No. 2, February 1997.

\bibitem{dolecek} G. Jovanovic-Dolecek and S.K. Mitra,
``A new two-stage sharpened comb decimator,'' {\em IEEE Trans. on
Circ. and Syst.-I,} Vol. 52, pp. 1414-1420, No. 7, July 2005.

\bibitem{dolecek_2} G. Jovanovic-Dolecek and S.K. Mitra,
``A new multistage comb-modified rotated sinc (RS) decimator with
sharpened magnitude response,'' {\em IEICE Transactions, Special
Issue on Recent Advances in Circuits and Systems,} Vol. 88-D, No.
7, pp. 1331-1339, July 2005.


\bibitem{laddomada} M. Laddomada and M. Mondin,
``Decimation schemes for $\Sigma\Delta$ A/D converters based on
Kaiser and Hamming sharpened filters,'' {\em IEE Proceedings of
Vision, Image and Signal Processing,} Vol. 151, No. 4, pp.
287-296, August 2004.

\bibitem{laddomada_sharp} M. Laddomada,
``Comb-based decimation filters for $\Sigma \Delta$ A/D
converters: Novel schemes and comparisons,'' {\em IEEE Trans. on
Sig. Proc.,} Vol.55, No. 5, Part 1, pp. 1769-1779, May 2007.

\bibitem{dolecek_3} G. Jovanovic-Dolecek and S.K. Mitra,
``A new two-stage CIC-based decimation filter,'' {\em In Proc. of
ISPA 2007,} pp. 218-223, Sept. 2007.

\bibitem{dolecek_4} G. Jovanovic-Dolecek and S.K. Mitra,
``On design of CIC decimation filter with improved response,''
{\em In Proc. of ISCCSP 2008,} pp. 1072-1076, 12-14 March 2008.

\bibitem{dolecek_5} G. Jovanovic-Dolecek,
`` A new modified comb-rotated sinc (RS) decimator with improved
magnitude response,'' {\em In Proc. of  ICECS 2007,} pp. 250-253,
11-14 Dec. 2007.

\bibitem{laddomada_part_poly} M. Laddomada,
``On the polyphase decomposition for design of generalized comb
decimation filters,'' {\em To appear on IEEE Trans. on Circuits
and Systems I,} 2008.

\bibitem{Diniz_book} P.S.R. Diniz, E.A.B. da Silva, and S. Lima Netto,
{\em Digital Signal Processing: System Analysis and Design},
Cambridge University Press, Cambridge, UK, 2002, ISBN
0-521-78175-2.

\bibitem{antoniou} A. Antoniou,
{\em Digital Signal Processing: Signals, Systems, and Filters},
McGraw-Hill, 2005, ISBN 0-07-145425-X.


\bibitem{Crochiere_StatApproach} R. Crochiere,
``A new statistical approach to the coefficient word length
problem for digital filters,'' {\em IEEE Trans. on Circuits and
Systems,} Vol.22, No. 3, pp. 190-196, March 1975.

\bibitem{Avenhaus} E. Avenhaus,
``On the design of digital filters with coefficients of limited
word length,'' {\em IEEE Trans. on Audio and Electroacoustics,}
Vol.20, No. 3, pp. 206-212, August 1972.

\bibitem{CrochiereOppenheim} R.E. Crochiere and A.V. Oppenheim,
``Analysis of linear digital networks,'' {\em Proceedings of the
IEEE,} Vol.63, No. 4, pp. 581-595, April 1975.


\end{thebibliography}
\end{document}